\documentclass[
aps,%
12pt,%
final,%
oneside,%
nobibnotes,%
nofootinbib,% 
superscriptaddress,%
centertags,
floatfix,
secnumarabic]%
{revtex4}%

\usepackage{graphicx}
\usepackage{amssymb,amsmath}
\usepackage{bm}
\RequirePackage[utf8]{inputenc}
\RequirePackage[T2A]{fontenc}
\RequirePackage[russian,english]{babel}
\usepackage{multirow}
\usepackage{verbatim}
\usepackage{wrapfig}
\usepackage{amsmath}
\usepackage{braket}
\newcommand{\Br}{\mathrm{Br}}

\begin{document}

%\large
\title{$B_c$ excitations at LHC experiments}
\author{\firstname{A.~V.}~\surname{Berezhnoy}}
\email{Alexander.Berezhnoy@cern.ch}
\affiliation{SINP MSU, Moscow, Russia}
\author{\firstname{I.~N.}~\surname{Belov}}
\email{in.belov@physics.msu.ru}
\affiliation{Phys. Dept. of MSU, Moscow, Russia}
\author{\firstname{A.~K.}~\surname{Likhoded}}
\email{Anatolii.Likhoded@ihep.ru}
\affiliation{NRC “Kurchatov Institute”–IHEP, Protvino, Moscow Region, 142281, Russia}
\affiliation{Moscow Institute of Physics and Technology}
\author{\firstname{A.~V.}~\surname{Luchinsky}}
\email{Alexey.Luchinsky@ihep.ru}
\affiliation{NRC “Kurchatov Institute”–IHEP, Protvino, Moscow Region, 142281, Russia}

\begin{abstract}
\normalsize
The  $2S$ $B_c$ states observed by ATLAS, CMS and LHCb Collaborations are discussed. 
The observation perspectives  of $B_c^*$, $2P$ wave, $3P$ wave and $D$ wave states of $B_c$ at LHC experiments are estimated. 

\end{abstract}

\maketitle

\section{Introduction}

$B_c$-meson is a nonrelativistic two heavy quark system, and like $(b\bar{b})$ and $(c\bar{c})$ states 
can be considered in the potential model framework that predicts 19 bounded states $(b\bar{c})$
below the decay threshold to $B\bar{D}$ (see \cite{Gershtein:1994jw,Gershtein:1997qy, Gouz:2002kk,Godfrey:2004ya}). 
Unlike $(b\bar{b})$- or $(c\bar{c})$-quarkonium, there are no annihilation decay modes for these states,
which makes the $(b\bar{c})$ system similar to usual $b\bar{q}$ or $c\bar{q}$-meson.
The first observation of the ground state $B_c$-meson was done at Tevatron collider (CDF and D0 experiments) 
in two decay modes: $B_c \to J\psi l \nu$ ($l = e, \mu$) and $B_c \to J\psi \pi$~\cite{Aaltonen:2007gv,Abazov:2008rba,Abazov:2008kv,Abe:1998wi}.  
Now this discovery is repeatedly confirmed by the LHC  Collaborations LHCb, CMS, ATLAS in numeral   decay modes: $B_c \to J\psi \pi$~\cite{Aaij:2012dd,CMS:2012oxa,ATLAS:2012bja}, $B_c \to J\psi \pi \pi \pi$~\cite{LHCb:2012ag,CMS:2012oxa}, $B_c \to J/\psi l\nu$~\cite{Aaij:2017tyk}, $B_c \to J/\psi K$~\cite{Aaij:2013vcx}, $B_c \to \psi(2S) \pi$~\cite{Aaij:2013oya}, $B_c \to J/\psi \pi K K$~\cite{Aaij:2013gxa}, $B_c \to J\psi D^{(*)} K^{(*)}$~\cite{Aaij:2016qlz, Aaij:2017kea}, $B_c \to J/\psi D^{(*)}_s$~\cite{Aaij:2013gia,Aad:2015eza}, $B_c \to J/\psi \pi p \bar p$~\cite{Aaij:2014asa}, $B_c \to J/\psi 3\pi^+ 2\pi^-$~\cite{Aaij:2014bla} and $B_c\to B_s\pi^+$~\cite{Aaij:2013cda}.  It is worth to note that most of these decays occur  due to transformation $b\to c W$.  The only observed mode with $c$ quark decay is $B_c\to B_s\pi^+$. The annihilation channel is not observed yet. 
It is interesting to note  that according to theoretical predictions the dominant decay channel is  $c\to s W$, the contribution of  $b\to c W$ is smaller, and the smallest  contribution is due to weak annihilation channel $bc\to W$. 

Also  the mass and life time are known with excellent accuracy~\cite{Aaij:2016qlz,Aaij:2014gka,Tanabashi:2018oca}:
$$M_{B_c}=6274.9 \pm 0.8 \mbox{ MeV},$$
$$\tau_{B_c}=0.507 \pm 0.009 \mbox{ ps}.$$

There are some differences in the production of ordinary quarkonium with hidden flavor and the production of $B_c$-meson. 
In case of $B_c$-meson it is necessary to produce two pairs of heavy quarks, $b\bar b$ and $c\bar c$, 
which leads to the strong suppression of the $B_c$ production cross section.  Unlike production of quarkonium with 
hidden flavor, the production of $P$-wave states is suppressed comparing to $S$-wave states. The production mechanism of $B_c$-meson and its excitations is well studied theoretically in~\cite{Berezhnoy:1994ba,Chang:1994aw, Berezhnoy:1995au,Kolodziej:1995nv,Berezhnoy:1996ks,Berezhnoy:1997fp,Baranov:1997sg, Baranov:1997wv, Berezhnoy:1997uz, Chang:2004bh, Berezhnoy:2004gc,Chang:2005wd,Chang:2006xka,Berezhnoy:2010wa,Gao:2010zzc}.   
The ratio $R_{B_c} = \sigma(B_c^*)/\sigma(B_c)$ is predicted to be about $\sim 2.6$ under assumption that the values of wave functions at origin for  $B_c^*$ and $B_c$  are the same. It is worth to note that according this study the $B_c$-meson production  at fixed gluon interaction energy resolves itself to just a $b$-quark fragmentation (like fragmentation $b \to B$) only at high transverse momenta, where the cross section ratio between the production of vector state $B_c^*$ and pseudoscalar state is predicted to be $R_{B_c}  \sim 1.4$.\footnote{It is worth to mention the study \cite{Zheng:2019gnb}, where the fragmentation functions $b\to B_c$ and $b\to B_c^*$ are estimated within one loop approximation and it is shown that value of  $R_{B_c}$ does not change dramatically. }  However the transverse momentum value where the fragmentation becomes dominant depends on the gluon energy: the higher gluon energy, the higher this transverse momentum value. This is why the non-fragmentation mechanisms contribute to all transverse momenta. 

  Unfortunately the observation of $B_c^*$ is extremely difficult due to the low energy of the photon in the decay $B_c^* \to B_c \gamma$. This is why in study~\cite{Berezhnoy:2013sla} we discussed  the other $B_c$ excitations, namely, $2S$-wave states and $P$-wave states, and these states are more preferable for observation. 
The recent observation of $2S$ states in ATLAS~\cite{Aad:2014laa},  CMS~\cite{Sirunyan:2019osb} and LHCb~\cite{Aaij:2019ldo} made us return to this  discussion.

Although the current state and prospects of research of $\bar b c$ quarkonia are discussed in an excellent review~\cite{Eichten:2019gig}, we nevertheless  would like to pay attention to some details.

The rest of the paper is organized as follows. In the next section decays of $B_c(2S)$ states with the emission of two $\pi$ mesons are discussed. Section \ref{sec:DWave} is devoted to the description of $D$-wave states' spectroscopy and production. In sections \ref{sec:radiative} and \ref{sec:BcLL} we give theoretical predictions for radiative decays of the excited states and lepton pair production in these processes. The results of our paper are summarized in the Conclusion.

\section{$B_c(2S)\to B_c(1S)+\pi\pi$ decays.}
\label{sec:2pi}

The excitations of $B_c$ meson were first observed by ATLAS~\cite{Aad:2014laa} and CMS~\cite{Sirunyan:2019osb} collaborations. Recently the results of CMS are confirmed by LHCb Collaboration in~\cite{Aaij:2019ldo}. The experimental results are shown in Table~\ref{tab:exp_Bc2S}.

\begin{table}[t]
\caption{Mass values and row relative yields of $B_c(2S)$ states estimated from the experimental data. }
\label{tab:exp_Bc2S} 
\begin{tabular}{||l|l|l|l|l||}
\hline
\multirow{2}{*}{}                   & experiment           & ATLAS~\cite{Aad:2014laa}      & CMS~\cite{Sirunyan:2019osb} & LHCb~\cite{Aaij:2019ldo}     \\ \cline{2-5} 
                                    & luminosity (energy) & 24.1 fb$^{-1}$ (7, 8 TeV)      & 140 fb$^{-1}$ (13 TeV)      & 8.7~fb$^{-1}$ (7, 8, 13 TeV) \\ \hline
\multirow{2}{*}{mass, MeV}          & $2^3S_1$ , shifted  & \multirow{2}{*}{$6842 \pm 6$} & $6842 \pm 2$                & $6841 \pm 1$                 \\ \cline{2-2} \cline{4-5}
                                    & $2^1S_0$            &                               & $6871.0 \pm 1.6$            & $6872.1 \pm 1.6$             \\ \hline
\multirow{3}{*}{row relative yield} & $2^3S_1$            & \multirow{2}{*}{}             & $0.0088\pm 0.0014$          & $0.0136 \pm 0.0027$          \\ \cline{2-2} \cline{4-5} 
                                    & $2^1S_0$            &                               & $0.0068 \pm 0.0014$         & $0.0063 \pm 0.0024$          \\ \cline{2-5} 
                                    & total               & $0.18 \pm 0.05$               & $0.0156 \pm 0.0019$         & $0.0198 \pm 0.0036$          \\ \hline
\multicolumn{2}{||c|}{$N(2^3S_1)/N(2^1S_0)$}              &                               & $1.31 \pm 0.32$            & $2.1 \pm 0.9$            \\ \hline
\end{tabular}
\end{table}

 %^1S_0 (B_c) &\xrightarrow[\sim 50 \%]{\pi^+\pi^-} 1 ^1S_0(B_c)\\

Before discussing these measurements, let us remind what was theoretically predicted. According to~\cite{Berezhnoy:2013sla} about one half of such excitations decay to $B_c$ ($B_c^*$) and $\pi^+\pi^-$ pair:
\begin{align*}
 2 ^1S_0 (B_c) &\xrightarrow[\sim 50 \%]{\pi^+\pi^-} 1 ^1S_0(B_c),\\
 2 ^3S_1 (B_c) &\xrightarrow[\sim 40 \%]{\pi^+\pi^-} 1 ^3S_1 (B_c),\\
\sigma(B_c(2S))&/\sigma^{\rm total}(B_c) \sim 25~\%.
\end{align*}
Therefore, the predicted relative  yield of $2S$-excitations with $B_c(2S) \to B_c(B_c^*)+\pi^+\pi^-$ decay is up to 10\% of total $B_c$-meson yield. Under assumption that the radial wave functions at origin  for $2 ^3S_1$ and for $2 ^1S_0$ states are approximately equal to each other, we predicted that   for the total phase space
\begin{align*}
\sigma(2 ^3S_1)/\sigma(2 ^1S_0 ) \sim 2.6.
\end{align*}
However, some models~\cite{Martynenko:2019Bc2Swf,Galkin:2019Bc2Swf, Ebert:2011jc}  predict that the wave function value is essentially larger for pseudoscalar $2S$ state,   than  for vector one.  According ~\cite{Martynenko:2019Bc2Swf} $|R(B_c^*(2S))(0)| /  |R(B_c(2S))(0)|=0.87$ that leads to  decreasing of $\sigma(2 ^3S_1)/\sigma(2 ^1S_0 ) $ from 2.6 to 2.  Within the approach~\cite{Galkin:2019Bc2Swf, Ebert:2011jc} the  $\sigma(2 ^3S_1)/\sigma(2 ^1S_0 ) $ decreases even more dramatically: $|R(B_c^*(2S))(0)| /  |R(B_c(2S))(0)|=0.567$ and therefore the ratio $\sigma(2 ^3S_1)/\sigma(2 ^1S_0 ) $ becomes close to 0.9.

It was also predicted in \cite{Berezhnoy:2013sla} that  the loss of ``soft'' photon from $B_c^*$ shifts the vector $2S$-state approximately by 65~MeV and insignificantly broadens  the peak:

\begin{equation}
\Delta \tilde{M}_{2S}  < 2 \frac{\Delta M^* \sqrt{\Delta M^2 - 4m_{\pi}^2}}{M}\approx 10~\textrm{MeV},
\label{eq:Delta_2S_max}
\end{equation}
where
$M$ is a ground state mass, $\Delta M^* = M(B_c^*)-M(B_c)$  is a difference between masses of lowest vector and pseudoscalar states and $\Delta M = M(B_c(2 ^3S_1))-M(B_c^*)$ is a difference between the masses of $2 ^3S_1$-wave excitation and lowest vector state. As a result, the mass peak on $B_c+\pi\pi$ spectrum for more massive vector state $2 ^3S_1$ will appear about 30~MeV lower, than for less massive pseudoscalar state $2 ^1S_0$. It is worth noting that in our previous study~\cite{Berezhnoy:2013sla} we have skipped the fact  that the $\Delta \tilde{M}_{2S} $ is  not a value of additional width itself,  but an upper value of additional width. The real value of additional width depends on shape of $\pi\pi$ distribution: the smaller the average value of $\pi\pi$ invariant mass $m_{\pi\pi}$, the narrower the peak. The width can be estimated as follows:
\begin{equation}
\Delta \tilde{M}_{2S}  \sim 2 \frac{\Delta M^* \langle \sqrt{\Delta M^2 - m_{\pi\pi}^2}\rangle}{M}.
\label{eq:Delta_2S}
\end{equation}

Now let us return to the experimental results (see Table~\ref{tab:exp_Bc2S}).
At first sight the measured $B_c(2S)$ yields are essentially smaller than the predicted one, which is about 10\%. However the registration efficiencies in these experimental studies were not taken into account and therefore at the moment this theoretical prediction does not contradict CMS and LHCb results. 

Also it is worth to note that the ATLAS measurement~\cite{Aad:2014laa} is significantly out of the range of others, and we think that such a huge difference can not be explained by the difference in the registration efficiencies.
 However the mass value measured by ATLAS is in consistence with shifted value of the vector $B_c(2S)$ excitation. The resolution in ATLAS is not enough to separate peaks and it makes awkward the detailed comparison with the results of other experiments, where the two peaks are clearly seen.
 
 It is very interesting to compare   the ratio of yields of $B_c^*(2S)$ and $B_c(2S)$  in CMS and LHCb experiments.  If LHCb measured the  central value which is close to the predicted one in~\cite{Berezhnoy:2013sla}: $\sim 2.1$, the CMS Collaboration   measured the central value $\sim 1.3$.  However, the errors are quite large, and within uncertainties these results do not contradict  each other. 
 
We hope that new data will allow to compare the ratio of yields
for different experiments. The situation would be most interesting if the ratio essentially depended on kinematic region, which would indicate a crucial change of production mechanism. 
It is worth noting that the sharp change of this value is not expected within the conventional production model.

\begin{figure}[!t]
\begin{center}
\begin{minipage}[h]{0.45\linewidth}
\includegraphics[width=1.2\linewidth]{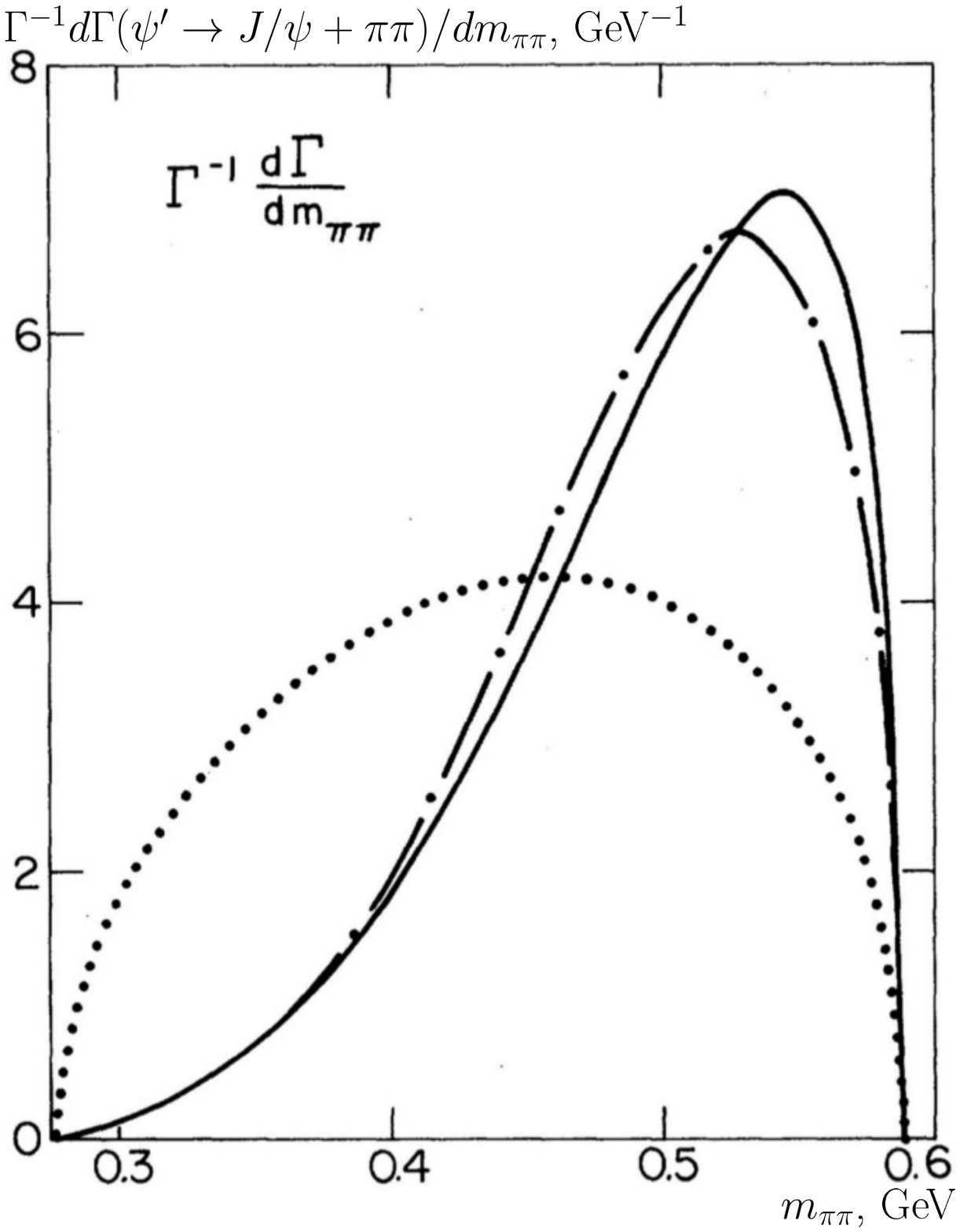}

\vspace*{-1.5cm}
\caption{The distribution over the invariant two pion mass $m_{\pi\pi}$ in the decay $\psi' \to J/\psi \pi\pi$ according  \cite{Brown:1975dz}: the solid and dot-dashed curves are obtained using the chiral symmetry  hypothesis without and with accounting of the interaction in the final state, correspondingly; the dotted curve is a phase space. (See also \cite{Novikov:1980fa,Voloshin:1975yb,Voloshin:1980zf}.)}
\label{fig:jpsipipi_old}
\end{minipage}
\hfill
\begin{minipage}[h]{0.45\linewidth}
\hspace*{-1.cm}
\vspace*{1.5cm}
\includegraphics[width=1.3\linewidth]{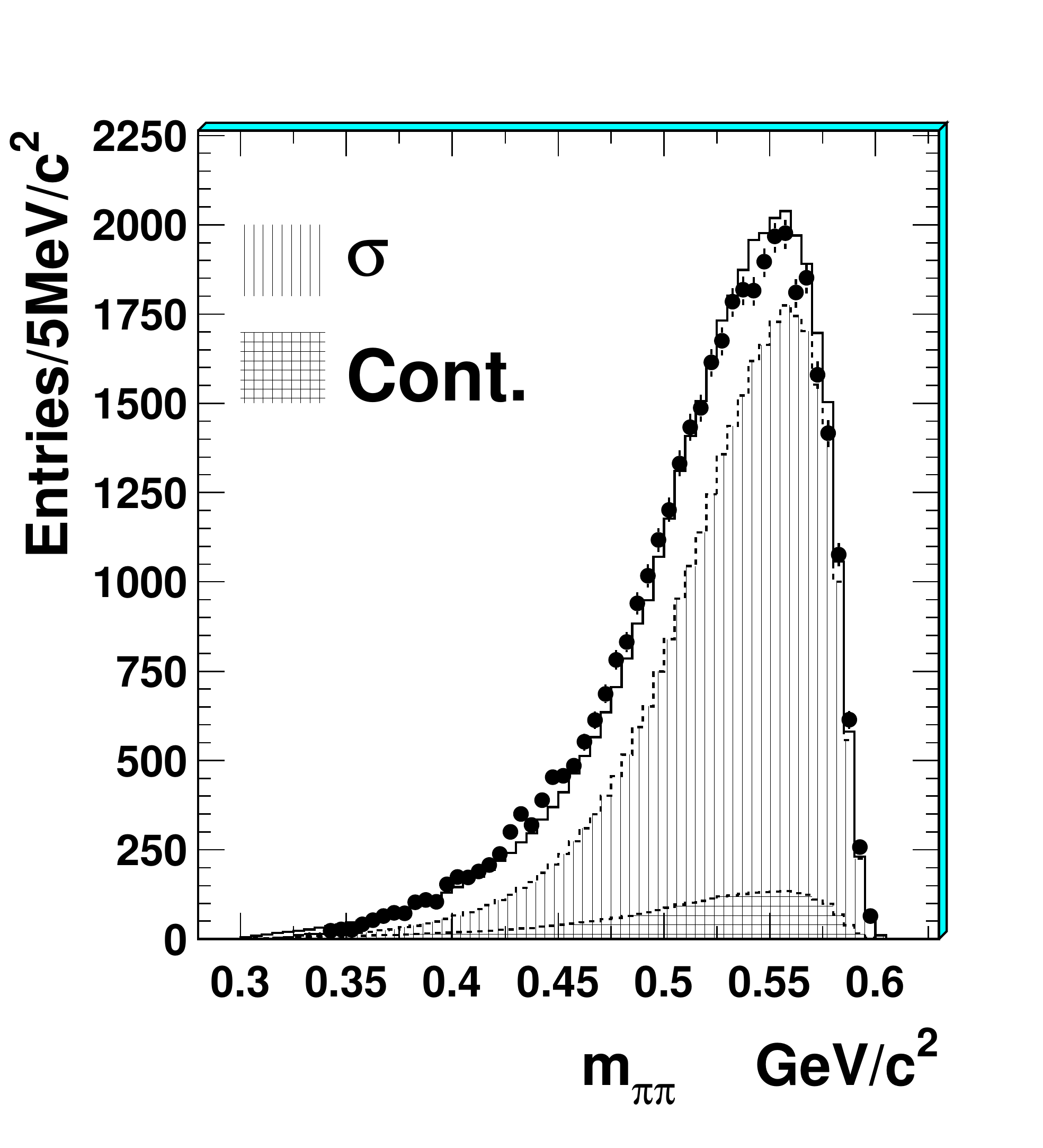}

\caption{The experimental distribution over invariant mass of two pions $m_{\pi\pi}$ in the process $\psi'\to J/\psi \pi\pi$ (BESII)~\cite{Ablikim:2006bz} The vertical hatching denotes a supposed  yield from $\sigma$ ($f_0(600)$) resonance.}
\label{fig:jpsipipi_exp}
\end{minipage}
\end{center}
\end{figure}

It will be very interesting to measure the shape of the spectrum of the invariant mass of $\pi\pi$-pair in this decay and compare it with similar spectrum 
shape for $\psi' \to \psi + \pi^+\pi^-$ decay. A theoretical distribution of the $\pi\pi$-pair mass in this decay has been investigated since 1970s. 
In~\cite{Brown:1975dz,Novikov:1980fa,Voloshin:1975yb,Voloshin:1980zf} from the considerations based on chiral theory it was concluded that for small invariant 
masses of $\pi\pi$-pair the amplitude of this process is approximately proportional to $m_{\pi\pi}^2-4m_{\pi}^2$. This approximation was extended to the whole 
phase space of the decay (see Figure~\ref{fig:jpsipipi_old}). However, 
the data from the BESII experiment~\cite{Ablikim:2006bz} shows, that it is very likely that resonant contribution to this process should be also considered. 
The resonant contribution is made by $f_0(600)$- or $\sigma$-resonance with $J^{PC}= 0^{+ +}$ and the mass $(400 \div 550)-i(200\div 350)$ MeV (Figure~\ref{fig:jpsipipi_exp}).

It should be noted that there is no consensus about the nature of the $\sigma$-meson. Currently, there are generally two approaches about it.
The first approach~\cite{Albaladejo:2012te} is based on unitary chiral perturbation theory, considering $\sigma$ as a dynamically generated resonance 
in $\pi\pi$ interactions. The $\sigma$-meson apprears to be an $S$-wave $\pi\pi$ bound state that can be also four-quark state.
Another approach~\cite{Black:2000qq, Harada:2012km} originates from linear sigma model. Within this approach, $\sigma$ is
a mixed state of two-quark and four-quark states. 
As seen from $\psi(2S)$ decay, the contribution from $\sigma$ dominates over raw $\pi\pi$-pair production. It is likely that $B_c(2S)$ meson will 
hadronically decay via the same or similar mechanisms, so the role of $\sigma$-meson in this case will be also dominant. The role of $\sigma$-meson in
$D_1(2430) \to D\pi\pi$ decay was studied in~paper \cite{Harada:2012km}.

\section{$D$-wave excitations at LHC}
\label{sec:DWave}

The production of $D$-wave states of $B_c$ quarkonium is not broadly discussed  due toupposedly small relative yield $\sim 1$\% (see for example~\cite{Cheung:1995ir}, where the production   of $D$-waves in $e^+e^-$ annihilation was studied), as well as  due to technical difficulties in the cross section estimation. However, at the moment, when  $2S$ states are already experimentally observed, the theoretical study of hadronic production of $D$-wave states looks quite reasonable.
Indeed, despite that the dominant decay mode for $D$-wave states is electromagnetic~\cite{Gouz:2002kk,Godfrey:2004ya,Ebert:2002pp,Li:2019tbn}, it is shown in~\cite{Eichten:1994gt} that about 20\% of such states decay to $1S$ state radiating two $\pi$ mesons, as well as $2S$ excitations. Thus it provides a chance to extract the $D$-wave states in the $B_c \pi^+\pi^-$ mass spectrum with large statistics. 

Similar to registered $B_c(2S) \xrightarrow{\pi\pi} B_c(1S)$ decay there should be several peaks: corresponding to direct decay to $B_c$ ground state and corresponding to decay with intermediate $B_c^*$, i.e. lowest vector state. The predicted $B_c$ spectroscopy comprises four $D$-wave states (we indicate their masses gained by different groups in table \eqref{tab:BcD}). If  $B_c(3D) \xrightarrow{\pi\pi} B_c(1S)$ decay goes with conservation of spin (as supposed in~\cite{Eichten:1994gt}), than we should obtain one peak for $3 ^1D_2$ state and three peaks for $3 ^3D_1, 3 ^3D_2, 3 ^3D_3$ states shifted by the value close to $\Delta M^* = M_{B_c^*} - M_{B_c}$. Most likely the latter three ones will overlap each other and will appear earlier than $3 ^1D_2$ peak at the $B_c \pi^+\pi^-$ invariant mass scale. Therefore we could expect, that  one narrow peak from $D$ wave states can be observed at $\sim$ 7000 MeV  and one broad peak can be observed at $\sim 6930$ MeV.

\begin{table}[ht]
\caption{Masses of $D$-wave $B_c$ meson states in MeV.}
\label{tab:BcD}
\begin{tabular}{||l|c|c|c|c|c|c|c|c|c||}
\hline
State & EQ~\cite{Eichten:1994gt} & GKLT~\cite{Gershtein:1994jw} & ZVR~\cite{Zeng:1994vj} & FUI~\cite{Fulcher:1998ka} & EFG~\cite{Ebert:2002pp} & GI~\cite{Godfrey:2004ya}  & MBV~\cite{Monteiro:2016ijw} & SJSCP~\cite{Soni:2017wvy} & LLLGZ~\cite{Li:2019tbn}\\
 \hline
 $3 ^3D_1$ & 7012 & 7008 & 7010 & 7024 & 7072 & 7028 & 6973 & 6998 & 7020\\
 $3 D_2'$ & \ldots & 7016 & \ldots & \ldots & 7079 & 7036 & 7003 & \ldots & 7032\\
 $3 D_2$ & \ldots & 7001 & \ldots & \ldots & 7077 & 7041 & 6974 & \ldots & 7024\\
 $3 ^1D_2$ & 7009 & \ldots & 7020 & 7023 & \ldots & \ldots & \ldots &  6994 & \ldots \\ 
 $3 ^3D_2$ & 7012 & \ldots & 7030 & 7025 & \ldots & \ldots & \ldots & 6997 & \ldots \\
 $3 ^3D_3$ & 7005 & 7007 & 7040 & 7022 & 7081 & 7045 & 7004 & 6990 & 7030\\
\hline
\end{tabular}
\end{table}

%%% Local Variables:
%%% mode: latex
%%% TeX-master: "Bc_excitations"
%%% End:

\section{Radiative $B_c$ meson decays}
\label{sec:radiative}

According to predictions of the potential model, the mass difference between lowest vector and pseudoscalar states of $\bar b c$-quarkonium ($B_c^*$ and $B_c$) is fairly small:
$$M(B_c^*)- M(B_c)\approx 65~\mbox{MeV}.$$
Since LHCb is unable to detect photon with transverse momentum about $65$~MeV it is necessary for decaying $B_c^*$ to have fairly large transverse momentum. 
It is known that production cross section is greatly reduced with increasing of the transverse momentum, so that it leads to significant decreasing of amount of events, 
where such photon could be detected.

The maximum photon transverse energy in the laboratory system can be calculated using the expression
\begin{equation}
\omega_T^{max}  \approx\Bigl( \sqrt{M_{B_c^*}^2+p_T^2}+p_T\Bigr)\frac{\Delta M^*}{M_{B_c^*}} 
\approx 0.01\Bigl( \sqrt{M_{B_c^*}^2+p_T^2}+p_T\Bigr) 
\label{busted_photon}
\end{equation}
This leads to, for example, that photons with transverse energy $\omega_T> 0.5$~GeV may be produced  only if transverse momentum of $B_c^*$-meson $p_T(B_c^*)> 24$~GeV. 
Such a cut on transverse momentum decreases  the yield of $B_c^*$-mesons approximately by two orders (see~\cite{Berezhnoy:2013sla}). It  is not so for radiation transitions of $2P$-wave states. In the latter case the transverse energy could be sufficiently large,  even if initial state of $B_c(2P)$-meson would have small momentum. Therefore, despite the fact that the yield 
of $2P$-excitations is about 6-20\% of total yield of $B_c$-meson~\cite{Berezhnoy:1996ks,Chang:2006xka}, it is much easier to register decays of $2P$-excitations.  As it was estimated  in~\cite{Berezhnoy:2013sla}, the yield of $2P$-excitations emitting photon with $\omega_T> 0.5$~GeV is $25\div 50$ times more, than yield of vector $B_c^*$, emitting photon with the same transverse energy.

It should be stressed that only 20\% of all $2P$-excitations emit only one photon, immediately transforming to lowest pseudoscalar state:
\begin{align*}
2P 1^+ (B_c)  &\xrightarrow[\sim 13 \%]{\gamma} 1 ^1S_0(B_c),\\
2P 1^{'+}(B_c) &\xrightarrow[\sim 94 \%]{\gamma} 1 ^1S_0 (B_c).
\end{align*}
In all other cases $2P$-states first decay to $B_c^*$ while emitting ``hard'' photon and then decay to $B_c$ emitting ``soft'' photon:
\begin{align*}
2P(B_c) \xrightarrow{\gamma^{\textrm{hard}}} {1 ^3S_1 (B_c^*)}  &\xrightarrow{\gamma^{\textrm{soft}}} {1 ^1S_0(B_c)}.
\end{align*}
The second photon (``soft'') will be lost almost always. However,  as shown below,  it doesn't prevent the experimental observation of $2P$-wave states of $B_c$-meson. 

\begin{table}
\caption{Radiative decays  of $B_c$ meson $P$-wave excitations (see~\cite{Godfrey:2004ya,Gupta:1995ps,Kiselev:1994rc})}
\begin{tabular}{||c|c|c|c||}
\hline
initial state & final state & Br, \% & $\Delta M$, MeV \\
\hline 
$2^3P_0$ & $1^3S_1+\gamma$ &  100 &  363-366 \\
$2 P 1^+$ & $1^3S_1+\gamma$ &  87 &  393-400 \\
 & $1^1S_0+\gamma$ &  13 &  393-400 \\
$2P 1'^+$ & $1^1S_0+\gamma$ &  94 &  472-476 \\
  & $1^3S_1+\gamma$ &  6  &  472-476 \\
$2^3P_2$ & $1^3S_1+\gamma$ &  100 &  410-426 \\
\hline
$3^3P_0$ & $1^3S_1+\gamma$ & 2  &  741 \\
$3 P 1^+$ & $1^3S_1+\gamma$ &  8.5 &  761 \\
 & $1^1S_0+\gamma$ &  3.3 &   820 \\
$3P 1'^+$ & $1^1S_0+\gamma$ &  22.6 &  825 \\
  & $1^3S_1+\gamma$ &  0.7 &  769 \\
$3^3P_2$ & $1^3S_1+\gamma$ &  18 &  778 \\

\hline
\end{tabular}
\label{tab:photon_decays}
\end{table}

Really, it is easy to show that the loss of the ``soft'' photon in the cascade decay of $2P$-wave states leads to broadening of the peak by the value 
\begin{equation}
\Delta\tilde{M} = \tilde{M}_{max}-\tilde{M}_{min}\approx 2 \frac{\Delta M^* \Delta M}{M}
\end{equation}
and to the left shifting it by $\Delta M^*$. Considering $\Delta M^*\approx 65$~MeV and $\Delta M\approx 400$~MeV we can get that the value of broadening for $2P$-wave states is
$$\Delta\tilde{M}_{2P} \approx 10~\textrm{MeV}.$$
It is clear that this width is smaller than hardware width of the resonance and doesn't affect its detection quality at all. 
It should be noted that the value of  broadening from ``soft'' photon loss for $3P$-wave states is also fairly small:
$$\Delta \tilde{M}_{3P} \approx 20~\textrm{MeV}.$$

\begin{figure}[!t]
\hspace{-2ex}
	\begin{minipage}[h]{0.49\linewidth}
		\center{\includegraphics[width=1.15\linewidth]{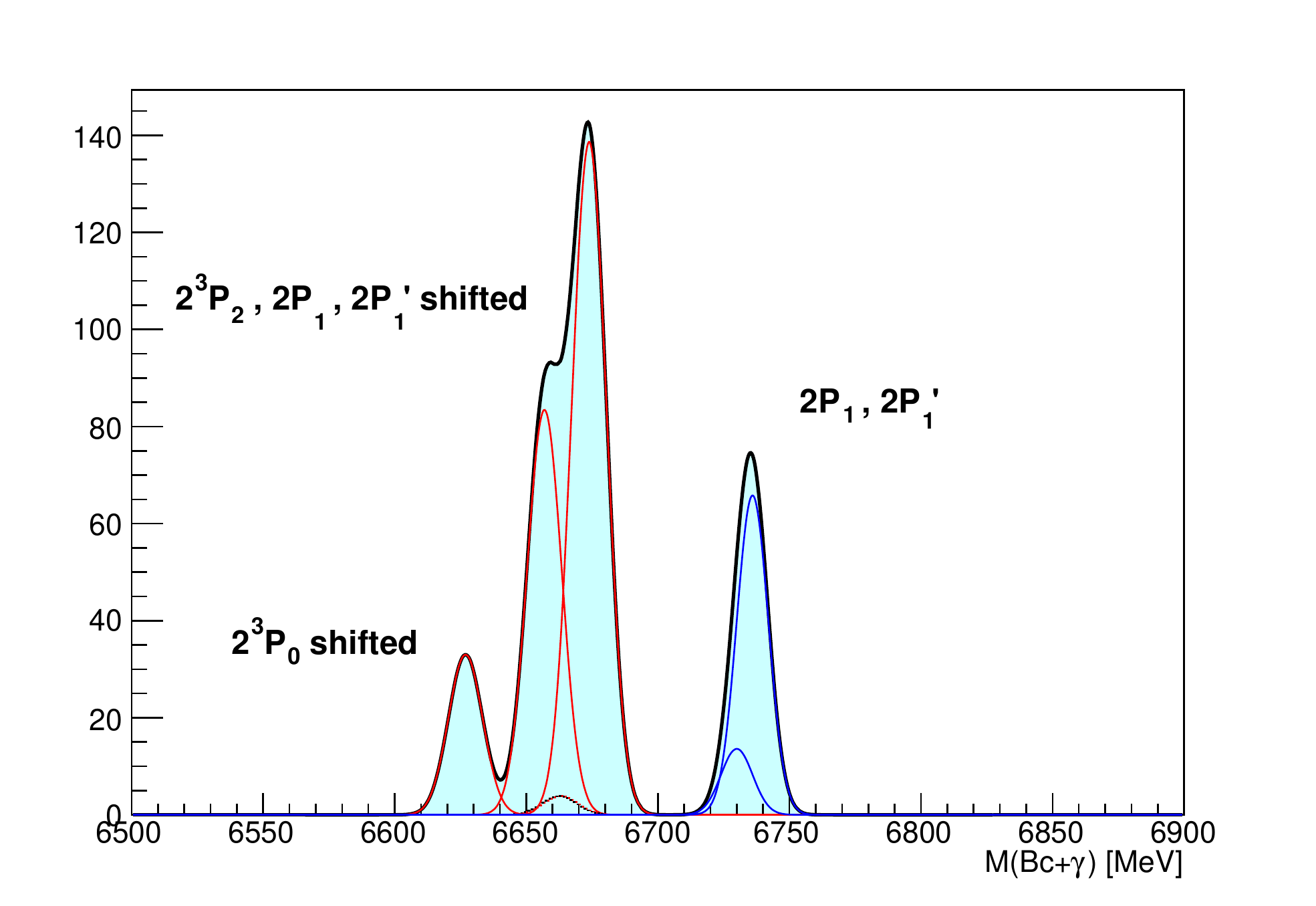}}\\
		(a)
	\end{minipage}
	\hfill
	\begin{minipage}[h]{0.49\linewidth}
		\center{\includegraphics[width=1.15\linewidth]{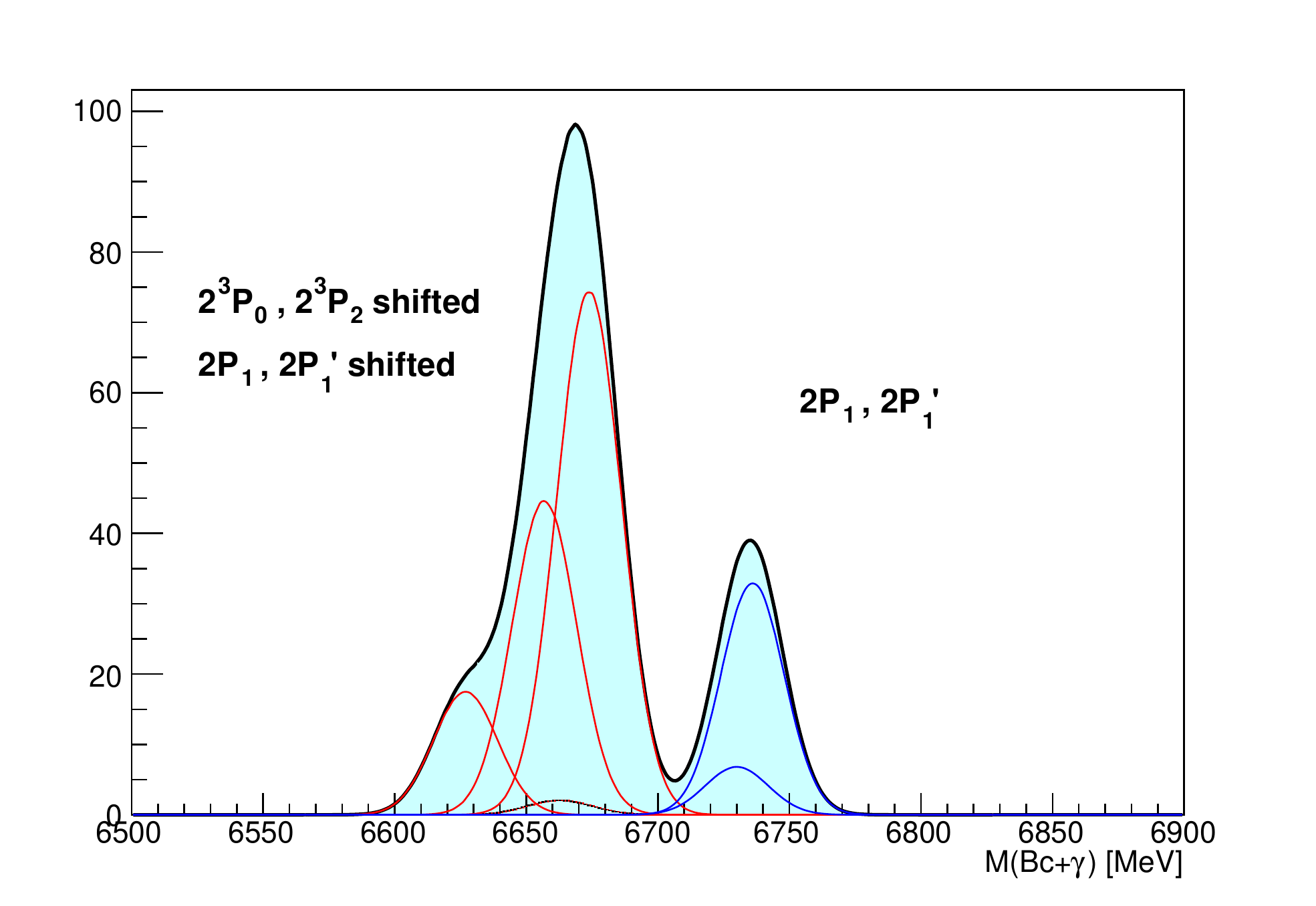}}\\
	   (b)
	\end{minipage}
\caption{  The mass spectrum of $B_c+\gamma^{\textrm{hard}}$ system  for the process $B_c(2P)\to B_c + \gamma^{\textrm{hard}} [+\gamma^{\textrm{soft}} ]$.}
\label{fig:2p}
\end{figure}
 
 \begin{figure}[!t]
\hspace{-2ex}
	\begin{minipage}[h]{0.49\linewidth}
		\center{\includegraphics[width=1.15\linewidth]{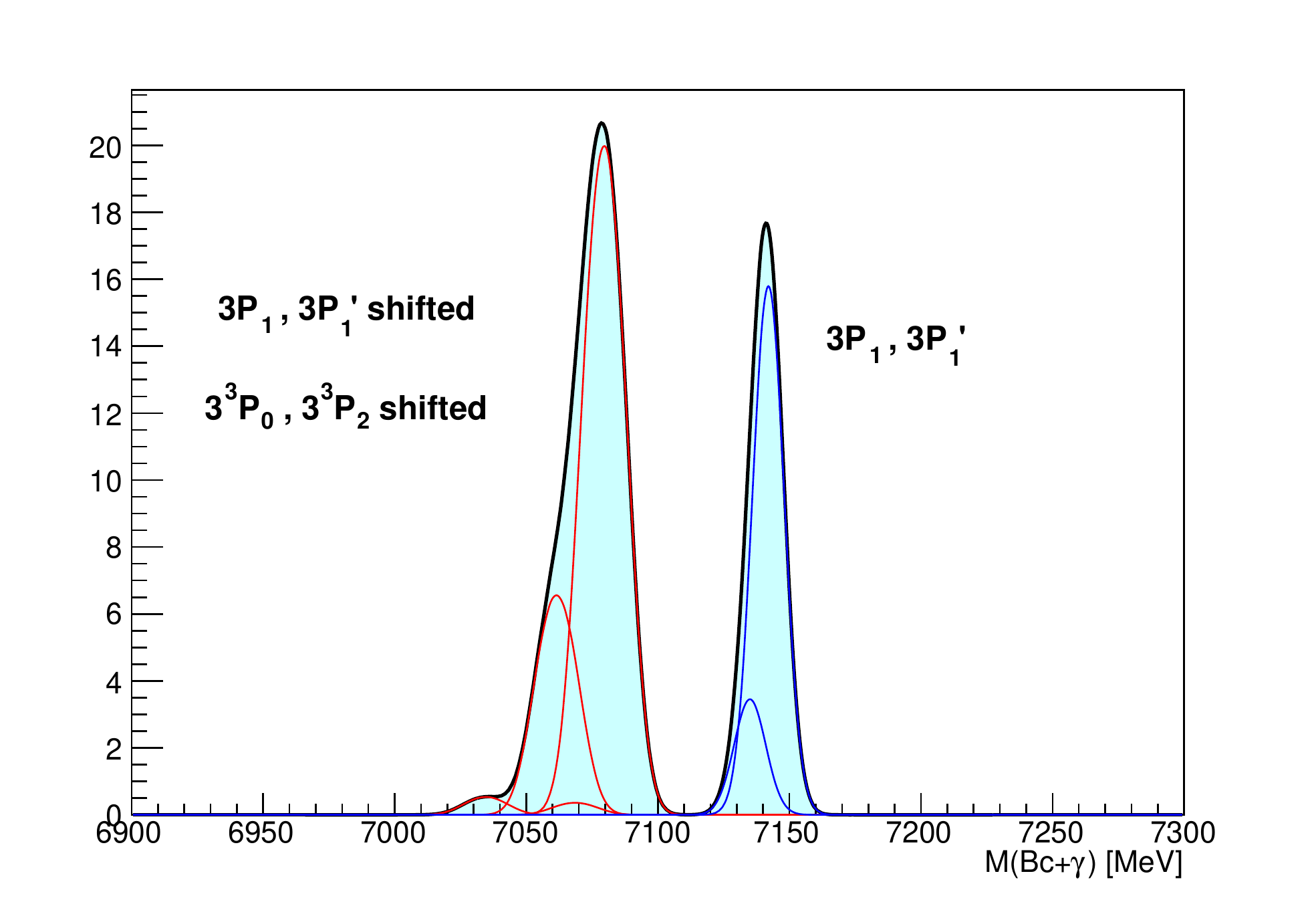}}\\
		(a)
	\end{minipage}
	\hfill
	\begin{minipage}[h]{0.49\linewidth}
		\center{\includegraphics[width=1.15\linewidth]{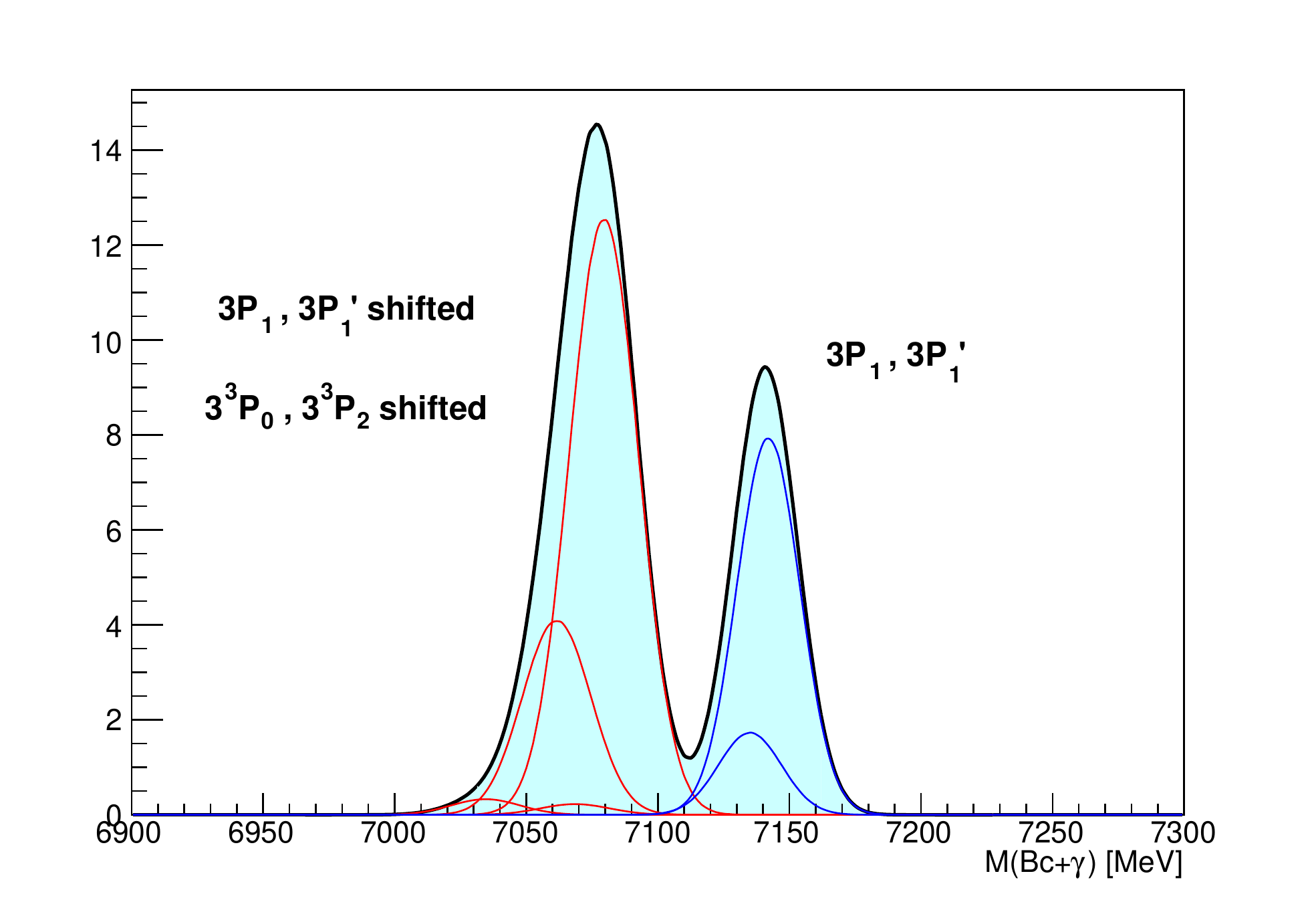}}\\
	   (b)
	\end{minipage}
\caption{  The mass spectrum of $B_c+\gamma^{\textrm{hard}}$ system  for the process $B_c(3P)\to B_c + \gamma^{\textrm{hard}} [+\gamma^{\textrm{soft}} ]$.}
\label{fig:3p}
\end{figure}

Predicted distributions over invariant mass of the final pseudoscalar $B_c$-meson and the ``hard'' photon in cascade decays of $2P$-wave states of the $B_c$-meson are shown in Figure~\ref{fig:2p}. Unshifted peaks from $2P~1^+$ and $2P~1^{'+}$ states are depicted with blue lines, red lines depict shifted and broadened peaks from $2^3P_0$, $2 P~1^+$, $2P~1'^+$ and $2^3P_2$ states. At last black line shows sum of the peaks. Similar distributions for $3P$-wave states are shown in Figure~\ref{fig:3p}. The pictures are plotted by the data from the table~\ref{tab:photon_decays} and $B_c$-meson mass spectrum from the work~\cite{Gouz:2002kk}. At all figures peaks are plotted with respect to rough  model of detector resolution: broadening the source histograms with Gaussian function with dispersion of $6$~MeV (pictures (a)) and $12$~MeV (pictures (b)). It should be stressed that such simulation is just an estimation and doesn't reflect the actual properties of detector. Nevertheless, such estimations allow us to understand what shapes would have peaks at finding them out in the experiment (see also~\cite{Eichten:2019gig}). 

It is important to note that in spite of the fact that the yields of $2P$- and $3P$-states in the proton-proton interaction are nearly the same, $3P$-excitations are harder to detect 
in the mass spectrum of $B_c+\gamma$. $2P$-excitations always decay via electromagnetic transitions, while $3P$-excitations --- only in 20\% of the cases. Moreover, it is clear that the shapes of  peaks broadened due to the loss of photon will repeat the shapes of distributions over the cosine of the angular between the directions of motion of ``soft'' and ``hard'' photons in rest frame of the decaying $B_c$ excitation.

Is is shown that the more minimal transverse energy of photon, the less probability it is radiated by $B_c^*$-meson.

\section{Lepton pair production in radiative $B_{c}$ meson decays}
\label{sec:BcLL}

The branching fraction of lepton pair production in radiative decays of the excited $B_{c}$ meson $B_{1}\to B_{2}\ell\ell$ can be written in the form \cite{Faessler:1999de,Luchinsky:2017pby}
\begin{align}
  \label{eq:br_MM}
  \frac{d\Br_{\ell\ell}}{dq^{2}} &= \frac{\alpha}{3\pi}\frac{1}{q^{2}}\frac{\lambda(M_{1};M_{2},\sqrt{q^{2}})}{\lambda(M_{1};M_{2},0)}\left(1-\frac{2m_{\ell}^{2}}{q^{2}}\right)\sqrt{1-\frac{4m_{\ell}^{2}}{q^{2}}}\Br_{\gamma}=\frac{dI^{\ell\ell}(q^{2})}{dq^{2}}\Br_{\gamma}
\end{align}
where $\Br_{\gamma}$ is the branching fraction of the original radiative decay, $q^{2}$ is the squared invariant mass of the lepton pair, $\alpha=e^{2}/4\pi$ is the fine structure coupling constant, $M_{1,2}$ are the masses of initial and final mesons respectively, and
\begin{align}
  \label{eq:lambda}
  \lambda(M;m_{1},m_{2}) &= \sqrt{1-\frac{(m_{1}+m_{2})^{2}}{M}} \sqrt{1-\frac{(m_{1}-m_{2})^{2}}{M}}
\end{align}
is the velocity of the final particles in $M\to m_{1}m_{2}$ decay.

It should be noted that the relation \eqref{eq:br_MM} is universal and does not depend on the physics of the process. The only assumption was that we neglect the $q^{2}$ dependence of the $B_{1}\to B_{2}\gamma^{*}$ decay vertex. This assumption looks quit reasonable since typical energy deposit in the radiative decays of the doubly heavy mesons is small in comparison with quarks' masses. As a result, the conversion factor $I^{\ell\ell}$ depends only on the masses of the initial and final particles. Masses of the leptons and ground state $B_{c}$ meson can be found easily \cite{Tanabashi:2018oca}, while for the initial excited particles some theoretical models are required.

In the following we will present numerical values of the $\gamma^{*}\to\ell\ell$ conversion factors for all the initial particles.

The spectroscopy of $B_{c}$ mesons was studied already in details. In our work we will use results presented in papers \cite{Godfrey:1985xj, Ebert:2002pp, Fulcher:1998ka, Kiselev:1994rc, Eichten:1994gt, Gupta:1995ps, Zeng:1994vj} (see also \cite{Godfrey:2004ya}). The mass of $B_{c}^{*}$ meson is not high enough for muon pair production, so only the $ee$ channel is opened. The corresponding results are shown in Table \ref{tab:conv_fact}. As for $P$ wave excitations, both electronic and muonic decays are allowed. One  can see in  Table \ref{tab:conv_fact}, that in all cases electron pair emission leads to suppression of the branching fraction by a factor $\sim 10^{-2}$, while in the case of $\mu\mu$ channel the suppression is about an order of magnitude stronger. However, unlike soft photon, the lepton-antilepton pair can be easily detected by the modern detectors, we think that the excited $B_{c}$ mesons could be observed in the discussed modes.

\begin{table}
  \centering
\begin{tabular}{||c|c|c|c|c|c|c|c||}
\hline
 & GI \cite{Godfrey:1985xj}  & EFG \cite{Ebert:2002pp}  & FUII \cite{Fulcher:1998ka}  & GKLT \cite{Kiselev:1994rc}  & EQ \cite{Eichten:1994gt}  & GJ \cite{Gupta:1995ps}  & ZVR \cite{Zeng:1994vj} \\ 
\hline
 mass of $B_c^*$, MeV  & $6.338$  & $6.332$  & $6.341$  & $6.317$  & $6.337$  & $6.308$  & $6.34$ \\ \hline
 $10^{3} \cdot I^{ee}$  & $6.105$  & $5.616$  & $5.431$  & $5.665$  & $5.869$  & $5.591$  & $6.011$ \\ 
\hline
\hline
 mass of $B_c(1P_1)$, MeV  & $6.741$  & $6.734$  & $6.737$  & $6.717$  & $6.73$  & $6.738$  & $6.73$ \\ \hline
 $10^{3} \cdot I^{ee}$  & $8.811$  & $8.733$  & $8.689$  & $8.733$  & $8.74$  & $8.821$  & $8.753$ \\  \hline
 $10^{3} \cdot I^{\mu\mu}$  & $0.7192$  & $0.6538$  & $0.6176$  & $0.6538$  & $0.6593$  & $0.7272$  & $0.6703$ \\ 
\hline
\hline
 mass of $B_c(1P_{1}')$,MeV & $6.75$  & $6.749$  & $6.76$  & $6.729$  & $6.736$  & $6.757$  & $6.74$ \\ 
 $10^{3} \cdot I^{ee}$  & $8.84$  & $8.782$  & $8.766$  & $8.773$  & $8.76$  & $8.88$  & $8.786$ \\ 
 $10^{3} \cdot I^{\mu\mu}$  & $0.7432$  & $0.6949$  & $0.6813$  & $0.6867$  & $0.6758$  & $0.7774$  & $0.6976$ \\ 
\hline\end{tabular}
  \caption{Conversion factors for $B_c^*,B_c(1P_1), B_c(1P_1)' \to B_c\gamma$ decays}
  \label{tab:conv_fact}
\end{table}

\section{Conclusion}
\label{sec:Conclusion}

In this work we briefly discuss the excellent results    of LHC Collaborations CMS, ATLAS and LHCb, which recently opened a new era in the heavy quark spectroscopy by  observing $2S$ excitations of $B_c$ meson.   We show that the study of $\pi^+\pi^-$  distribution could provide a new information about the nature of $\sigma$-meson and about the chiral symmetry breaking. We emphasize  that the ratio dependence  between yields of $2 ^3S_1$ and $2 ^1S_0$ states of $B_c$ meson on the kinematical conditions is the very  important source of information about the $B_c$ meson production mechanisms.
Also we estimate the perspectives of observation of $B_c$ states, such as  $B_c^*$, $P$ wave and $D$ wave excitations. We think, that at least for $P$ excitations such prespectives are fairly optimistic. In addition we suggest to study $P$ wave states in their radiative decays to the lepton pair.

Authors thank V. Galkin and A. Martynenko for help and fruitful discussion.

A. Berezhnoy and I. Belov  acknowledge the support from  ``Basis'' Foundation (grants 17-12-244-1 and 17-12-244-41).

\clearpage
\appendix

\section{Maximum $p_T$ of the photon emitted by the excited $B_c$}
We consider single-photon transition $B_c^* \xrightarrow{\gamma}{B_c}$.
In the $B_c^*$ rest frame photon energy $\omega'$ can be determined from the equation:
$$
M_{B_c^*} = \omega' + \sqrt{\omega'^{\ 2} + M_{B_c}^2}  
$$
\begin{equation}\label{omega}
\omega' = \Delta M^* \left(1-\frac{\Delta M^*}{2M_{B_c^*}}\right),
\end{equation}
where $\Delta M^*$ is a mass difference between $B_c$ states: $\Delta M^* = M_{B_c^*} - M_{B_c}$. 

The energy $\omega$  of the photon in the laboratory system achieves maximum,  if the direction of the photon motion  coincides with the direction of the $B_c^*$ motion in this system:
\begin{equation}
\omega^{max}=\frac{(1+v)\omega'}{\sqrt{1-v^2}} =  \frac{\omega'}{M_{B_c^*}}\left(\sqrt{p_{B_c^*} + M_{B_c^*}^2} + p_{B_c^*} \right),
\end{equation}
where $v=\frac{p_{B_c^*}}{E_{B_c^*}}$.

The maximal transverse energy for the given  values of $p_{B_c^*}$ and $p_{B_c^*} ^T$ can be estimated as
$$
\omega^{max }_{T} (p_{B_c^*}^T,p_{B_c^*})=\omega^{max} \frac{p_{B_c^*}^T}{p_{B_c^*}} =  \frac{\omega'}{M_{B_c^*}}\left(\sqrt{p_{B_c^*}^T + M_{B_c^*}^2\Bigl(\frac{p_{B_c^*}^T}{p_{B_c^*}}\Bigr)^2} + p_{B_c^*}^T. \right)
$$

For the maximal transverse energy for the given  value of $p_{B_c^*}^T$ and arbitrary $p_{B_c^*}$ value we get in terms of $p_{B_c^*}^T$:
\begin{equation}\label{wtmax}
\omega_T^{max} = \left(1 - \frac{\Delta M^*}{2M_{B_c^*}}\right) \Bigl( \sqrt{M_{B_c^*}^2+{p_{B_c^*}^T}^2} + p_{B_c^*}^T\Bigr)\frac{\Delta M^*}{M_{B_c^*}}.
\end{equation}

Expression \eqref{wtmax} is valid for any electromagnetic transition. Particularly for the considered case (lowest vector state decaying into the ground state):
$$
\omega_T^{max} \approx 0.01 \Bigl( \sqrt{M_{B_c^*}^2+{p_{B_c^*}^T}^2} + p_T\Bigr)
$$

\section{The additional decay width due to the loss of soft photon in the decay  $B_c(2 ^3S_1) \xrightarrow{\pi\pi} {B_c(1 ^3S_1)}  \xrightarrow{\gamma}{B_c(1 ^1S_0)}.$}

We denote masses of $B_c$ states as:
\begin{align*}
&B_c(2 ^3S_1):\ \ \ M + \Delta M^* + \Delta M \\
&B_c(1 ^3S_1):\ \ \ M + \Delta M^* \\
&B_c(1 ^1S_0):\ \ \ M
\end{align*}

Invariant mass squared for the initial and final states:
\begin{equation}\label{two}
\left(M + \Delta M^* + \Delta M\right)^2 = \left(M + \varepsilon + \omega\right)^2 - \left(k + p\right)^2, 
\end{equation}
where $(\varepsilon, p)$ and $(\omega, k)$ are four-momenta of $\pi\pi$ and $\gamma$ correspondingly in the $B_c$ ground state rest frame. 

Invariant mass squared for the decay $B_c(1 ^3S_1) \rightarrow B_c(1 ^1S_0) + \gamma:$
\begin{equation}\label{one}
\left(M + \Delta M^*\right)^2 = \left(M + \omega\right)^2 - k^2.
\end{equation}
$$
\omega = \Delta M^*\left(1 + \frac{\Delta M^*}{2 M}\right)
$$

Solving \eqref{two} by $\varepsilon$ we get:
\begin{align*}\notag
\varepsilon_{max} &= \frac{1}{M\left(M+2\omega\right)}\left[\left(M+\omega\right)F + \omega\sqrt{F^2 - M m_{\pi\pi}^2 \left(M+2\omega\right)} \right], \\
\varepsilon_{min} &= \frac{1}{M\left(M+2\omega\right)}\left[\left(M+\omega\right)F - \omega\sqrt{F^2 - M m_{\pi\pi}^2 \left(M+2\omega\right)} \right], 
\end{align*}
where $F = \frac{1}{2}\left(\Delta M^2 + 2 \Delta M\left(\Delta M^* +  M\right) - m_{\pi\pi}^2\right)$

Further we write approximately with $\frac{m_{\pi\pi}}{M},\frac{\Delta M^*}{M},\frac{\Delta M^*}{M}$ treated as very small.
$$
\varepsilon_{min}^{max} \approx \Delta M \left(1 - \frac{\Delta M^*}{M}\right) \pm \Delta M^*\sqrt{\left(\frac{\Delta M}{M}\right)^2 - \left(\frac{m_{\pi\pi}}{M}\right)^2}
$$

Invariant mass of reсonstructed $B_c(1 ^1S_0) + \pi\pi$ system: 
$$
\tilde{M}_{2S} = \sqrt{\left(M + \varepsilon\right)^2 - p^2} 
$$
Taking $\tilde{M}_{2S}$ with $\varepsilon_{min}, \varepsilon_{max}$ values we can write it in the following form:
\begin{equation}
\tilde{M}_{2S} \approx M\left(1 + \frac{\Delta M}{M} + \ldots \right) \pm \frac{\Delta\tilde{M}_{2S}}{2},
\end{equation}
where we do not specify the terms of higher order contributing to $\tilde{M}_{2S}$ shift from the exact value $M_{2S} = M + \Delta M + \Delta M^*$. Thus we obtain the extra broadening
\begin{equation}
\Delta\tilde{M}_{2S} \approx 2\Delta M^*\sqrt{\left(\frac{\Delta M}{M}\right)^2 - \left(\frac{m_{\pi\pi}}{M}\right)^2}. 
\end{equation}

\section{The additional decay width due to the loss of soft photon in the decay $B_c(2 ^3P_1) \xrightarrow{\gamma^{\textrm{hard}}} {B_c(1 ^3S_1)}  \xrightarrow{\gamma^{\textrm{soft}}} {B_c(1 ^1S_0)}.$}

We denote masses of $B_c$ states as:
\begin{align*}
&B_c(2 ^3P_1):\ \ \ M + \Delta M^* + \Delta M \\
&B_c(1 ^3S_1):\ \ \ M + \Delta M^* \\
&B_c(1 ^1S_0):\ \ \ M
\end{align*}

Invariant mass squared for the initial and final states:
\begin{equation}\label{eone}
 (M+\Delta M +\Delta M^*)^2 = (M + \omega + \omega^*)^2 - (k + k^*)^2,
 \end{equation}
 where $(\omega, k)$ and $(\omega^*, k^*)$ -- four momenta of $\gamma^{\textrm{hard}}$ and $\gamma^{\textrm{soft}}$ in the $B_c$ ground state rest frame and $\cos\Theta$ is an angle between $k$ and $k^*$.

Invariant mass squared for the decay $B_c(1 ^3S_1) \rightarrow B_c(1 ^1S_0) + \gamma^{\textrm{soft}}:$
\begin{equation}\label{etwo}
(M + \Delta M^*)^2 = (M + \omega^*)^2 - k^{*2}
\end{equation}

From \eqref{eone} and \eqref{etwo} we get:
\begin{align*}
&\omega^* = \left(1+\frac{\Delta M^*}{2M}\right) \Delta M^* \\
&\omega = \frac{2M+\Delta M + 2\Delta M^* }{2M+2\omega^* (1-\cos\Theta)} \Delta M
\end{align*}
Approximately:
$$
\omega \approx \left(1+\frac{\Delta M}{2M} + \frac{\Delta M^*}{M} \cos\Theta \right) \Delta M\ ,
$$
where $\frac{\Delta M}{M}$ and $\frac{\Delta M^*}{M}$ are very small.

We consider that $\gamma^{\textrm{soft}}$ is being lost in the reconstructed invariant mass: 
\begin{equation}\label{ethree}
\tilde{M}_{2P} = \sqrt{\left(M + \omega\right)^2 - k^2} \approx M + \omega
\end{equation}
We obtain that 
$$
\tilde{M}_{2P} \sim \frac{\Delta M \Delta M^*}{M}\cos\Theta\ ,
$$  
and $\tilde{M}_{2P}$ peak gains extra broadening with varying $\Theta$: 
\begin{equation}
\Delta\tilde{M}_{2P} \approx 2\frac{\Delta M \Delta M^*}{M} 
\end{equation}

\bibliography{Bc_excitations}

\end{document}